\documentclass[12pt,preprint]{aastex}
\def\etal{{\it et al.\ }}
\begin{document}
\title{Generation of Flows in the Solar Chromosphere Due to 
Magnetofluid Coupling}
\author{Swadesh M. Mahajan\altaffilmark{1}}
\affil{Institute for Fusion Studies, The University of Texas at  Austin, 
Austin, Texas 78712}
\author{Komunela I. Nikol'skaya\altaffilmark{2}}
\affil{Institute of Terrestrial Magnetism, Ionosphere and Radio Wave 
Propagation, Troitsk of Moscow Region, 142092, Russia}
\author{Nana L. Shatashvili\altaffilmark{\,3,0} and Zensho 
Yoshida\altaffilmark{4}}
\affil{Graduate School of Frontier Sciences and High Temperature 
Plasma Center, The University of Tokyo, Tokyo 113--0033, Japan}
\altaffiltext{0}{\small {\it Permanent address:} \ Plasma Physics 
Department, Tbilisi State University, Tbilisi 380028, Georgia} 
\altaffiltext{1}{\small Electronic mail: \ mahajan@mail.utexas.edu}
\altaffiltext{2}{\small Electronic mail: \ knikol@izmiran.troitsk.ru}
\altaffiltext{3}{\small Electronic mail: \ 
nana@plasma.q.t.u-tokyo.ac.jp\hskip 0.3cm nanas@iberiapac.ge}
\altaffiltext{4}{\small Electronic mail: \ yoshida@k.u-tokyo.ac.jp}

\clearpage

\begin{abstract}
 
It is shown that a generalized magneto--Bernoulli mechanism can 
effectively generate high velocity flows in the Solar chromosphere by 
transforming the plasma pressure energy into kinetic energy. It is 
found that at reasonable heights and for realistic plasma parameters, 
there is a precipitous pressure fall accompanied by  a sharp 
amplification of the flow speed.  
\end{abstract}


\clearpage

\section{Introduction}
\label{sec:intro} 
\bigskip

Recent observations, strongly fortified by immensely improved measuring 
and interpretive capabilities, have convincingly demonstrated that the 
solar corona is a highly dynamic arena replete with multiple--scale 
spatiotemporal structures (Aschwanden \etal 2001a).  A major new advance is the 
discovery that strong flows are found everywhere --- in the subcoronal 
(chromosphere) as well as in the coronal  regions  (see e.g. 
(Schrijver \etal 1999; Winebarger, LeLuca and Golub 2001; Wilhelm 
2001; Aschwanden \etal 2001a; Aschwanden \etal 2001b; Seaton \etal 
2001; Winebarger \etal 2002) and references therein). Equally important is the growing
belief and realization that the plasma flows may complement 
the abilities of the magnetic field in the creation of the 
amazing richness observed in the coronal structures.
 
Interestingly enough, even before the observational mandate, theoretical 
efforts in harnessing the plasma flows to solve some of the riddles of 
solar physics had already begun. In particular the dynamics of flow--based 
structure creation and heating was the subject of Mahajan \etal (1999, 2001); 
in this model the flows provided the basic material as well as  energy for 
the primary heating of the coronal loops. A systematic treatment of loop models
that include flows was also developed by Orlando, Peres and Serio (1995a, 1995b), 
and by Mahajan \etal (1999, 2001).

If flows are to play an important and essential role in determining the 
dynamics and structure of the solar corona, we must immediately face the 
problem of finding sources and mechanisms for the creation of these flows. 
Catastrophic models of flow production in which the magnetic energy is 
suddenly converted into bulk kinetic energy (and thermal energy) are 
rather well--known; various forms of magnetic reconnection (flares, micro 
and nanoflares) schemes permeate the literature (see e.g.~(Wilhelm 
2001; Christopoilou, Georgakilas and Koutchmy 2001)
for chromosphere up--flow generations). A few 
other  mechanism of this genre also exist: Uchida \etal (2001) proposed 
that the major part of the supply of energy and mass to the active regions 
of the corona may come from a dynamical leakage of magnetic twists produced in 
the subphotospheric convection layer; Ohsaki, Shatashvili, Yoshida and Mahajan (2001,2002) 
have shown how a slowly evolving closed structure 
(modelled as a double--Beltrami two--fluid equilibrium)  may experience, 
under appropriate conditions,  a sudden loss of equilibrium with the 
initial magnetic energy appearing as the mass flow energy. Another   
mechanism, based on loop interactions and fragmentations and  explaining 
the formation of loop threads, was given in Sakai and Furusawa (2002). These 
mechanisms, though extremely interesting, are an unlikely source to account 
for  the observed ubiquity of plasma flows.  One should, perhaps, look for 
relatively gentler, more widespread, and  steadier mechanisms.

Before exploring new avenues for flow generation, and deciding which 
region of the solar atmosphere should be the first target for 
investigation, we seek some guidance from phenomenology. Based on 
estimates of energy fluxes required to heat the chromosphere and the 
corona, Goodman (2001) has shown  that the  mechanism which transports 
mechanical energy from the convection zone to the chromosphere 
(to sustain its heating rate) could also supply the energy needed 
to heat the corona, and accelerate the solar wind (SW). The coronal
heating problem, in this context, is shifted to the problem of the
dynamic energization of the chromosphere. In the latter process the 
role of flows is found to be critical as warranted by the following 
observations made in soft X--rays and extreme ultraviolet (EUV) wavelengths, 
and recent findings from the {\it Transition Region and Coronal Explorer 
(TRACE)}: the overdensity of coronal loops, the chromospheric up--flows 
of heated plasma, and the localization of the heating function in the 
lower corona (e.g. Schrijver, \etal 1999; Aschwanden \etal 2001a; 
Aschwanden 2001b) and references therein). The
main message then, is that the coronal heating problem may only be solved
by including processes (including the flow dynamics) in the chromosphere 
and the transition region. The challenge, therefore, is to develop a 
semi--steady state theory of flow generation in the chromosphere.

There are only two obvious energy sources that could power flow generation 
in the chromosphere: the magnetic field, and the thermal pressure of the 
plasma. We have already mentioned a few examples of the magnetically 
driven transient but sudden flow--generation. Looking for a more quiescent pathway, 
we shall now concentrate on the latter source. To convert thermal energy into kinetic 
energy, some variant of the Bernoulli mechanism (the existence of larger kinetic energy
in regions of lower pressure) must be invoked. We will soon show that  the 
double--Beltrami--Bernoulli states accessible to a two--fluid  system in 
which the velocity field is formally treated at par with the magnetic 
field (Mahajan and Yoshida 1998), can readily provide the necessary framework for this conversion.

\section{Model}
\label{sec:Model}

In astrophysics (particularly  in the physics of the solar 
atmosphere), it is useful to assign at least two distinct connotations
to the plasma ``flow'': 1)~the flow is a primary object whose dynamics bears 
critically on the phenomena under investigation. The problems of the formation 
and the original heating of the coronal structure, the creation of channels for 
particle escape, for instance, fall in this category, 2)~The flow is a secondary 
feature of the system, possibly created as a by-product (e.g. see 
Ohsaki \etal (2002)) and/or used to drive or suppress an instability. 
Since the generation of flows 
which will eventually create the coronal loops is the theme of this effort, the 
flows  here are  fundamental. 

As pointed out earlier, our theoretical model is based on a rather simple 
application of the magnetofluid theory developed over the last few 
years. We plan to restrict ourselves to almost steady state considerations 
(for a steady and continuous supply of plasma flows emerging from  the 
chromosphere). Very near the photospheric surface, the influence of 
neutrals and ionization processes (and processes of flux emergence etc.) 
would not permit a quasi--equilibrium approach. A little farther distance 
($\Delta r$) from the surface, however, we expect that the quasi--equilibrium 
two--fluid model will capture the essential physics of flow generation. 

>From recent observational data (see e.g. Goodman (2000); Aschwanden 
\etal (2001a); Socas-Navarro and Almeida (2002) and 
references therein) we could obtain the following average plasma density 
and temperature at $\Delta r\sim (500-2000)\,$km: $n\sim 
(10^{15}-10^{11})\,{\rm cm}^{-3}; \ T\sim 1\,$eV (for simplicity we will assume equal 
electron and ion temperatures). The information about the magnetic field 
is a little harder to extract because of the low sensitivity and lack of 
high spatial resolution of the measurements coupled with the inhomogeneity 
and co--existence of small-- and large--scale structures with 
different temperatures in  nearby regions. At these distances we have 
different values for the network and for the internetwork fields. The network 
plasmas have typically short--scale fields in the range $B_0\sim (700-1500)\,$G, 
have more or less uniform density and will be prone to explosive/eruptive 
analysis of the kind carried out in Ohsaki \etal (2002). 
The internetwork fields, on the other hand, are generally smaller (with 
some exceptions (Socas-Navarro and Almeida (2002)) --- $B_o\sim 500\,$G, and are embedded in 
larger--scale plasma structures with inhomogeneous densities. The theory 
of steady creation of flows in the lower chromosphere  will be based on 
these latter objects.

\section{Generation of flows in lower chromosphere}
\label{sec:flows}

To illustrate the basic physics of flow creation, we deal with the 
simplest two--fluid equilibria with $T=\rm const$ leading to \ 
$n^{-1}\nabla p \to \,T\nabla \,\ln\,n$. The generalization to a homentropic 
fluid with  $p=\rm const\cdot n^{\gamma}$ is straightforward and is done in the numerical work.

The dimensionless equations describing the model equilibrium can be read 
off from Mahajan \etal (2001)
\begin{equation}
\frac{1}{n}\nabla\times {\bf b\times b}+\nabla\left(\frac{r_{A0}}{r}
-\beta_0\ln\ n-\frac{V^2}{2}\right)+ {\bf V\times (\nabla\times
V})=0, \label{eq:DB-1}
\end{equation}
\begin{equation}
\nabla\times\left({\bf V}-\frac{\alpha_0}{n}\nabla \times {\bf b}
\right)\times {\bf b}=0, \label{eq:DB-2}
\end{equation}
\begin{equation}
\nabla \cdot (n{\bf V})=0, \label{eq:cont}
\end{equation}
\begin{equation}
\nabla\cdot{\bf b}=0, \label{eq:b}
\end{equation}
where the notation is standard with the following normalizations: the 
density $n$ to $n_0$ prevalent at about $500\,$km  (and farther) from the 
solar surface (this is where the plasma is created), the magnetic field to 
the ambient field strength at the same distance, and velocities to the 
Alfv\'en velocity $V_{A0}$. The parameters $r_{A0}=GM_\odot/V_{A0}^2R_{\odot}=2
\beta_0/r_{c0}, \ 
\alpha_0=\lambda_{i0}/R_{\odot}, \ \beta_0=c_{s0}^2/V_{A0}^2$ are defined 
with $n_0, \ T_0, \ B_0$. Here $c_{s0}$ is a sound speed, $R_\odot$ is the 
solar radius, $r_c=GM_\odot/2c_{s0}^2R_{\odot}$, and 
$\lambda_{i0}=c/\omega_{i0}$ is the collisionless skin depth.  

The double--Beltrami solutions, expressing the simple physics that the 
electrons follow the field lines, while the ions, due to their inertia, 
follow the field lines modified by the fluid vorticity, are contained in 
the pair,  
\begin{equation}
{\bf b}+\alpha_0 \nabla\times {\bf V}=d\ n\ {\bf V}, \qquad \qquad
{\bf b}=a\ n\ \left[{\bf V}-\frac{\alpha_0}{n}\,\nabla\times {\bf
b}\right], \label{eq:DB-3} 
\end{equation}
where $a$ and $d$ are dimensionless constants related to the two ideal
invariants, the magnetic and the generalized helicities (Mahajan and 
Yoshida 1998; Mahajan \etal 2001),
\begin{equation}
h_1=\int ({\bf A}\cdot {\bf b})\ d^3x , \label{eq:h1}
\end{equation}
\begin{equation}
h_2=\int ({\bf A}+{\bf V})\cdot ({\bf b}+\nabla\times {\bf V})d^3x. 
\label{eq:h2}
\end{equation}
On substituting (6)--(7) into (1)--(2), one obtains the Bernoulli condition
\begin{equation}
\nabla\left(\frac{2\beta_0r_{c0}}{r}-\beta_0\ln\,n-\frac{V^2}{2}\right)=0,
\label{eq:bernoulli}
\end{equation}
relating the density with the flow kinetic energy, and solar gravity. 
Equations~(\ref{eq:DB-1}), (\ref{eq:DB-3}), and (\ref{eq:bernoulli}) 
represent a close system, and may be easily manipulated to yield~($g(r)=r_{c0}/r$)
\begin{equation}
\frac{\alpha_0^2}{n}\nabla\times\nabla\times{\bf V}+\alpha_0\
\nabla\times \left[\left(\frac{1}{a\,n}-d\right)n\,{\bf 
V}\right]+\left(1-\frac{d}{a}\right){\bf V}=0, \label{eq:CC-1}
\end{equation}
\begin{equation}
\alpha_0^2\nabla\times\left(\frac{1}{n}\nabla\times{\bf 
b}\right)+\alpha_0\ \nabla \times \left[\left(\frac{1}{a\,n}-d\right){\bf 
b}\right]+\left(1-\frac{d}{a}\right){\bf b}=0,\label{eq:CC-2}
\end{equation}
\begin{equation}
n=\exp\left(-\left[2g_0-\frac{V^2_0}{2\beta_0}-2g+\frac{V^2}{2\beta_0}
\right]\right), \label{eq:density}
\end{equation}
a set ready to be solved for the density, the velocity and the magnetic 
field. We must point out that this time--independent set is not suitable 
for studying chromospheric heating processes (primary heating at lower 
heights). The main thrust of this paper is to uncover mechanisms which 
create flows in the  chromosphere --- flows that will supply matter 
and energy  needed to create the coronal structures, and provide their 
primary heating. The creation and heating problem, of course, requires a 
fully time--dependent treatment (Mahajan \etal 2001).

We have to resort to numerical methods to obtain detailed solutions for 
the coupled nonlinear system~(\ref{eq:CC-1}), (\ref{eq:CC-2}), and 
(\ref{eq:density}). We have carried out a 1D simulation (the relevant 
dimension being the height ``Z" from the center of the Sun; $Z_0=R_{\odot}+\Delta r$
is the surface at which the boundary conditions are applied) for a variety of 
boundary conditions. The boundary surface is so chosen that at this height $Z_0$ 
the ionization is expected to be complete. For estimates of $\Delta r$ given 
earlier, the relevant heights lie within 
$[(1+0.7\cdot 10^{-3})-(1+2.8\cdot 10^{-3})]\,R_{\odot}$.

The simulation results are presented in Figs.~1--2. These are the plots of 
various physical quantities as  functions of the height. The first figure 
consists of three frames (a--b, c--d, and e--f) each consisting of two 
pictures --- one for the density and the magnetic field and the other for 
the velocity field. The parameters defining  different frames are: 
$n_0\sim 10^{14}{\rm cm}^{-3}$, \ $B_0\sim 500\,$G, \  $V_{A0}\sim 110\,$km/s implying  
$\beta_0 \sim 0.01\ll 1$ and $r_{c0}=900$ for a--b; $n_0\sim 10^{16}{\rm cm}^{-3}$, 
\ $B_0\sim 1000\,$G, \ $V_{A0}\sim 22\,$km/s implying $\beta_0\sim 0.2< 1$ and 
$r_{c0}=950$ for c--d; and $n_0\sim 10^{17}{\rm cm}^{-3}$ and $B_0\sim 
1500\,$G, \ $V_{A0}\sim 11\,$km/s, \ $\beta_0 \sim 1$ and $r_{c0}=1000$ for the frame e--f. 
In each frame there are three sets of curves labelled by $\alpha_0$ (1--2--3 
corresponding respectively to $\alpha_0 =10^{-5},\, 10^{-3},\, 10^{-1}$), the 
measure of the strength of the two--fluid Hall currents.

For all our runs the temperature at the boundary was taken to be $\sim 
1\,$eV ($c_{s0}\sim 10\,$km/s), and the boundary conditions, $|b_0|=1, 
\quad V_0=1\,$km/s (with $V_{x0}=V_{y0}=V_{z0}$) were imposed. Notice that we 
begin with just a small residual flow speed. The choice, \ $d\sim a \sim 
100$ \  and \ $(a-d)/a^2\sim 10^{-6}$ \  for the parameters characterizing the 
double Beltrami state, reflects the physical constraint that we are 
dealing with a sub--Alfv\'enic flow with a very small $\alpha_0$ (Mahajan 
\etal 1999). We must admit that the values of $\alpha_0$ chosen for the 
simulation are much  larger than their actual values ($\sim 10^{-8}$ for 
corona and $\sim 10^{-11}$ for chromosphere); our present code cannot resolve 
the equivalent short lengths, though, we hope to do better in future. We believe, 
however, that the nature of the final results is properly captured by these 
artificial values of $\alpha_0$.

The most remarkable result of the simulation is that for small and 
realistic values of $\alpha_0$ (curves labelled 1), there exists some 
height where the density begins to drop precipitously  with a 
corresponding sharp rise in the flow speed. The effect is even stronger 
for the low beta (a--b are the lowest beta frames) plasmas. It is also 
obvious that at very short distances, the stratification is practically 
due to gravity, but as we approach the velocity ``blow--up'' height, the 
self--consistent magneto--Bernoulli processes take over and control the 
density (and hence the velocity) stratification.

An examination of the Bernoulli condition~(\ref{eq:density}) readily 
yields an indirect estimate for the height at which the observed 
shock formation may take place. For a low--beta plasma, the sharp fall in 
density is expected  to occur where the local flow kinetic energy exceeds 
the kinetic energy specified at the  boundary (this is true for all  $\alpha_0$), i.e, 
\begin{equation}
|{\bf V}|^2- V_0^2 > 2\,\beta_0. \label{eq:criter}
\end{equation}
For the current simulation, at  $\beta_0 =0.01$, it occurs approximately 
at $|{\bf V}|^2 > 0.02$ or at $|V|\sim 0.14$. This analytically--predicted 
value is very close to the simulation result (see Fig.~2(b)). Simulation 
results also confirm that the velocity blow--up distance depends mainly on 
$\beta_0$, and that the final velocity is greater for greater $\beta_0$ 
(Fig.~2(a)). The data presented in Fig.~1  and Fig.~2 corresponds to a uniform 
temperature plasma. For this case, the variations in plasma pressure are 
entirely due to the variations in density. Since the magnetic energy remains
practically uniform over the distance, sharp decrease in density with a 
corresponding sharp rise in the flow--speed (flow energy changes are of the 
order of $n^{-1/2}$) is nothing but the expression of the commonly understood 
Bernoulli effect. We must emphasize that the general results remain unchanged  
in our extensive simulations in which the temperature is allowed to vary 
(but we have to use a homentropic equation of state to analytically derive 
the Beltrami states. The final parameters, naturally, depend upon the 
adiabaticity index $\gamma $).  

Taking into account the fact that the fiducial height $Z_0$ is different 
for different cases (larger for smaller $\beta_0$), one expects  that the 
plots for the ``blow--up'' distance will approximately match one other, 
and there exists some minimum distance (possibly of the order $500\,$km or so) 
from the photosphere below which steady up--flows can not be generated. This 
expectation is simply supported by the numerical results given in  
Fig.~2(a). Observations clearly indicate that there exists a narrow layer 
above the photosphere where no spicules, mottles etc. are seen.

To check whether the generated flows are predominantly radial or somewhat 
more isotropic (to explain the observational constraints) we studied in 
detail the relatively large $\beta_0$ case (fixing $\beta_0$ is quite 
difficult due to complications like ionization) and found that the flows 
tend to be mostly radial only for large  $\alpha_0$ (see, for example, 
plots labeled 2 and 3 in Fig.~1(b,d,f)). The situation could change 
considerably when we deal with a more inclusive time--dependent dynamical 
model with dissipation. Plasma heating, then, could result from the 
dissipation of the perpendicular energy so that at larger distances, the 
flows would have larger radial components. Heating would also keep 
$\beta({\bf r},t)$ large at upper heights shifting the velocity blow--up 
distance further or eliminating it all together; we know from Fig.~1 that 
as $\beta_0$ goes up, the density fall(velocity amplification) becomes  
smoother. These issues will be dealt with later in a more detailed work.

Note, that if one were to ignore the flow term in (\ref{eq:bernoulli}) (a 
totally wrong assumption commonly used in many studies), we will end up 
finding essentially radial flows. The magnitude of these flows, however, 
remains small; there is no region of sharp rise (\ref{eq:criter}), and 
the  generated flows achieve reasonable energies at heights typically 10 
times greater than the heights at which the correct Bernoulli condition 
would do the trick. 

\section{Conclusions and Summary}
\label{sec:conclusions} 

We have shown a  possible pathway  for a steady generation of flows in the 
quasi--equilibrium stage established through the interaction of the emerging
magnetic fluxes with the existing cold solar chromosphere (when an ionized   
$\sim 1\,$eV plasma is trapped in internetwork structures). The suggested 
mechanism is a straightforward application of the recently--developed  
magnetofluid model (Mahajan and Yoshida 1998; Mahajan \etal 1999, 
2001); a generalized Bernoulli mechanism
(a necessary condition for the double--Beltrami magnetofluid equilibrium) 
allows the pressure energy to be very effectively transformed to flow kinetic energy 
as the plasma moves away from the sun. We find that at reasonable heights 
and for realistic plasma parameters, there is a precipitous pressure fall 
with a sharp amplification of the flow speed. In the presence of 
dissipation, these flows are likely to play a fundamental role in the 
heating of the inner and upper chromosphere, although our explicit purpose 
in this paper was to create  a steady source of matter and energy for the 
formation and primary heating of the corona. Our preliminary results agree 
with the observation data, and lend promise to attempts, based on the 
existence of subcoronal flows, to tackle unresolved problems like the 
coronal heating and origin of the solar wind. 
 
\section*{ACKNOWLEDGEMENTS}


Authors acknowledge the help of Dr. K.I. Sigua for preparing the 
simulation  pictures and thank Dr. E. Marsch for interesting and useful discussions.
All the authors thank Abdus Salam International Centre for Theoretical Physics, 
Trieste, Italy. The work of SMM was supported by USDOE Contract No.~DE--FG03--96ER--54366. 
Work of KIN was supported by a Grant No.~02--02--16199a 
of Russia Fund of Fundamental Research 
(RFFR).  The work of NLS was supported by ISTC Project G--633.

\clearpage

\figcaption[f1.eps]{Plots of density, magnetic fields and velocity 
versus height for values of $\alpha_0$ and $\beta_0$. Sub--figures (a) and 
(b) are for $\beta_0=0.01, \ r_{c0}=900$; (c) and (d) are for 
$\beta_0=0.2, \ r_{c0}=950$; (e) and (f) are for $\beta_0 = 1, \ 
r_{c0}=1000$. The numbers 1, 2, 3 represent  $\alpha_0 =10^{-5},\, 
10^{-3}\, 10^{-1}$ respectively. $V_y$ is not displayed since its 
behavior is practically similar to  $V_x$. The velocity blow--up is 
controlled by $\beta_0$. For a bigger (unrealistic)  $\alpha_0$ there is a 
splitting of the velocity components --- at the end the radial component 
is dominant. Magnetic field energy does not change much on these distances.}

\figcaption[f2.eps]{The ``blow--up" distance (a) and velocity (b)
versus $\alpha_0$. The smaller the $\beta_0$, the smaller the   
``blow--up" distance and smaller the velocity at ``blow--up" 
(compare with (\ref{eq:criter})). For fixed $\beta_0$, the process is less 
sensitive to changes in $\alpha_0$.}


\clearpage

\begin{thebibliography}{99}
    
 \bibitem[Aschwanden {\it et al.} 2001a]{A1} Aschwanden, M.J., Poland 
A.I., and Rabin D.M. 2001a, Ann. Rev. Astron. Astrophys., 39, 175.

\bibitem[Aschwanden 2001b]{A2} Aschwanden, M.J. 2001b, \apj, 560, 
1035. 

\bibitem[Christopoilou, Georgakilas, and Koutchmy 2001]{koutchmy} 
Christopoulou, E.B., Georgakilas, A.A., and Koutchmy, S. 2001, Solar 
Phys., 199, 61.

\bibitem[Goodman 2001]{goodman} Goodman, M.L. 2001, Space Sci. Rev., 
95, 79.

\bibitem[Goodman 2000]{goodman2} Goodman, M.L. 2000, \apj, 533, 501.

\bibitem[Mahajan {\it et al.} 1999]{MMNS1} Mahajan, S.M., Miklaszewski, 
R., Nikol'skaya, K.I., and Shatashvili N.L., February 1999, Preprint 
IFSR \#857, {\it The University of Texas, Austin}, 67 pages. 

\bibitem[Mahajan {\it et al.} 2001]{MMNS2}  Mahajan, S.M., Miklaszewski, 
R., Nikol'skaya, K.I., and Shatashvili, N.L. 2001, Phys. Plasmas, 8, 
1340. 

\bibitem[Mahajan and Yoshida 1998]{MY-1} Mahajan, S.M., and Yoshida, 
Z. 1998, Phys. Rev. Lett., 81, 4863. 

\bibitem[Nikol'skaya and Valchuk 1998]{nv} Nikol'skaya, K.I., and 
Valchuk, T.E. 1998, Geomagnetizm and Aeronomy, 38, No.~2, 14.

\bibitem[Ohsaki {\it et al.} 2001]{osym1} Ohsaki, S., Shatashvili, N.L., 
Yoshida, Z., and Mahajan, S.M. 2001, \apj, 559, L61.

\bibitem[Ohsaki {\it et al.} 2002]{osym2} Ohsaki, S., Shatashvili, N.L., 
Yoshida, Z., and Mahajan, S.M. 2002, \apj, 570.

\bibitem[Orlando, Peres and Serio 1995a]{flows1} Orlando, S., Peres, 
G., and Serio, S. 1995a, Astrophys. and Astron., 294, 861.

\bibitem[Orlando, Peres and Serio 1995b]{flows2} Orlando, S., Peres, 
G., and Serio, S. 1995b, Astrophys. and Astron., 300, 549.

\bibitem[Sakai and Furusawa 2002]{sakai1} Sakai, J.I., and Furusawa, 
K. 2002, \apj, 564, 1048.

\bibitem[Schrijver {\it et al.} 1999]{schrijver}  Schrijver, C.J., 
Title, A.M., Berger, T.E., Fletcher, L., Hurlburt, N.E., Nightingale, 
R.W., Shine, R.A., Tarbell, T.D., Wolfson, J., Golub, L., Bookbinder, 
J.A., Deluca, E.E., McMullen, R.A., Warren, H.P., Kankelborg, C.C., 
Handy B.N., and DePontieu, B. 1999, Solar Phys., 187, 261. 

\bibitem[Seaton {\it et al.} 2001]{ami1} Seaton, D.B., Winebarger, A.R., 
DeLuca, E.E., Golub, L., and Reeves, K.K. 2001, \apj, 563, L173.

\bibitem[Socas--Navarro and Sanchez Almeida 2002]{sa} Socas--Navarro, 
H., and Sanchez Almeida, J. 2002, \apj, 565, 1323.

\bibitem[Uchida {\it et al.} 2001]{uchida} Uchida Y., Miyagoshi, T., Yabiku 
T., Cable S., and Hirose S. 2001, Publ. Astron. Soc. Japan, 53, 331. 

\bibitem[Wilhelm 2001]{wilhelm} Wilhelm, K. 2001, Astrophys. and 
Astronomy, 360, 351.

\bibitem[Winebarger, DeLuca and Golub 2001]{golub} Winebarger, A.M., 
DeLuca, E.E., and Golub, L. 2001, \apj, 553, L81.

\bibitem[Winebarger {\it et al.} 2002]{ami2} Winebarger, A.R., Warren, 
H., Van Ballagooijen, A., DeLuca E.E., and Golub, L. 2002, \apj, 567, 
L89.

\end{thebibliography}
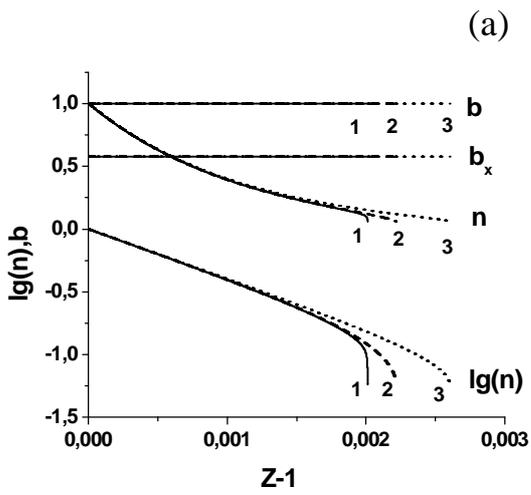
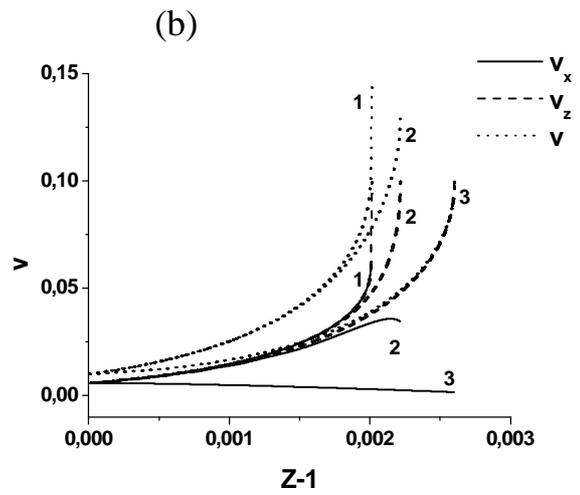
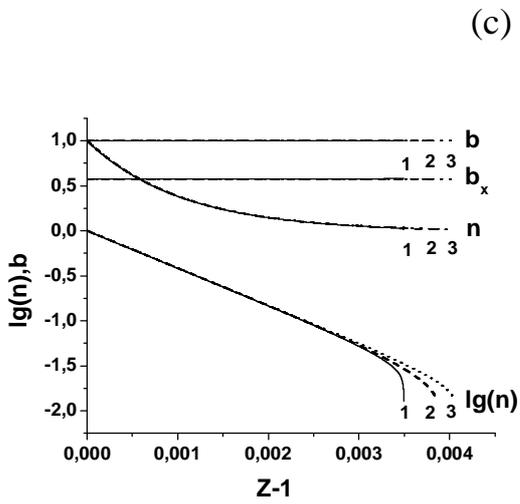
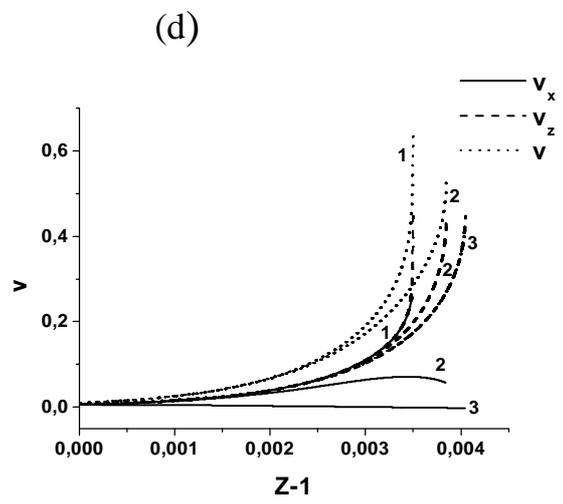
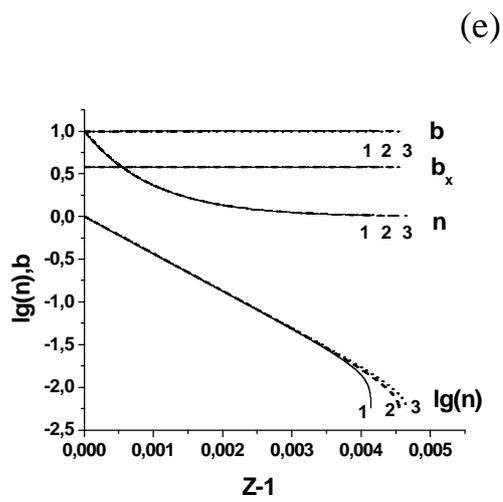
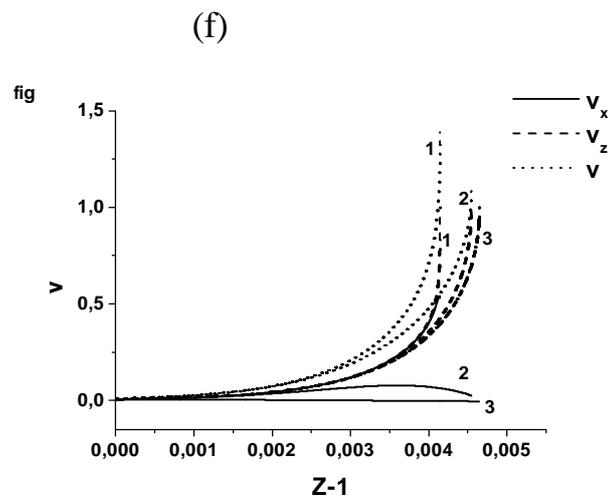
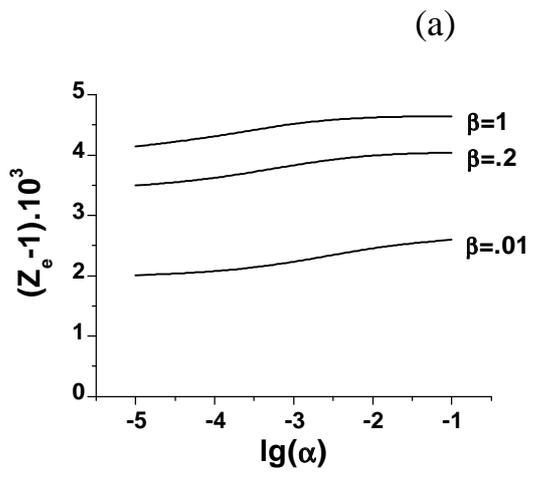
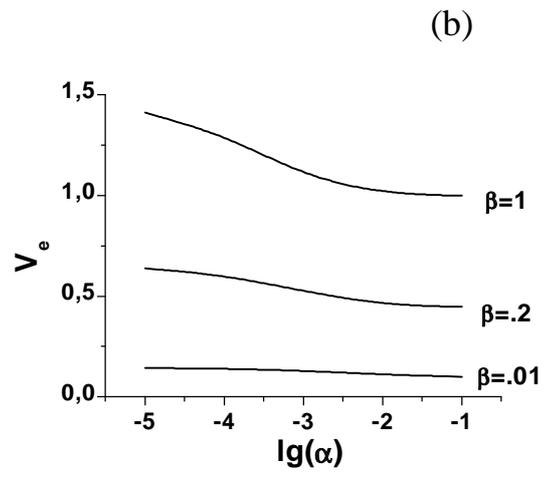
\end{document}